\documentclass[11pt]{article}
\usepackage{moriond,epsfig}
\usepackage[latin1]{inputenc}
\usepackage[OT1]{fontenc}

\def\be{\begin{equation}}
\def\ee{\end{equation}}
\def\bea{\begin{eqnarray}}
\def\eea{\end{eqnarray}}

\begin{document}
\title{HOLONOMY CORRECTIONS TO THE COSMOLOGICAL PRIMORDIAL TENSOR POWER SPECTRUM}

\author{A. BARRAU \& J. GRAIN}

\address{Laboratoire de Physique Subatomique et de Cosmologie,
UJF, CNRS-IN2P3-INPG\\
Laboratoire AstroParticule et Cosmologie, Universit\'e Paris 7, CNRS-IN2P3}

\maketitle\abstracts{
Loop quantum gravity is one of the leading candidate theory to non-perturbatively 
quantize gravity. In this framework, holonomy corrections to the equation of 
propagation of gravitons in a FLRW background have been derived. We investigate 
the consequences of those corrections on the tensor power spectrum in de-Sitter and
slow-roll inflations, for n=-1/2. Depending on the value of the Barbero-Immirzi parameter, several 
observational features could be expected. 
}

This brief note aims at providing a first estimate of possible Loop Quantum Gravity
footprints on the tensor power spectrum. It focuses on the specific n=-1/2 case
and the full study will be reported in a dedicated article \cite{grain}.\\

In canonical quantum gravity, dynamics is determined by a Hamiltonian (constraint)
operator rather than a path integral. Relevant quantum corrections can therefore
be obtained at the level of an effective Hamiltonian (as opposed to an effective
action in a covariant quantization). To study the quantum effects expected from
Loop Quantum Gravity (LGQ) corrections, an effective
tensor mode Hamiltonian has been derived by Bojowald \& Hossain \cite{bojo1}
(for a general introduction to LQG and Ashtekar variables, one can refer to the book
"Quantum Gravity" by Rovelli \cite{carlo}). In a canonical triad formulation of general relativity, there are 
three types of constraints: Gauss constraints (which generate local rotations of
the triads), diffeomorphism constraints (which generate spatial diffeomorphisms) 
and Hamiltonian constraints (which completes the space-time diffeomorphisms).
Tensor mode perturbations are completely governed by the Hamiltonian constraint.\\

Two basic types of quantum corrections are expected from the Hamiltonian of loop
quantum gravity. These corrections arise for inverse powers of the densitized
triad and from the fact that loop quantization is based on holonomies, {\it i.e.}
exponentials of the connection, rather than direct connection components. In the
following article we will only focus on holonomy corrections, neglecting inverse
volume effects and backreaction effects. Those points will be studied in another
paper \cite{grain}.\\

It was shown \cite{bojo1} that the propagation of gravitons in a flat FLRW 
background is given, when holonomy corrections (which provide higher order and higher spatial
derivative terms) are taken into account, by the 
following equation of motion~:

\begin{equation}
	\left[\frac{\partial^2}{\partial\eta^2}+\left(\frac{\sin{(2\gamma\bar\mu\bar{k})}}{\gamma\bar\mu}\right)\frac{\partial}{\partial\eta}-\nabla^2-2\gamma^2{\bar\mu}^2\left(\frac{\bar{p}}{\bar\mu}\frac{\partial\bar\mu}{\partial\bar{p}}\right)\left(\frac{\sin{(\gamma\bar\mu\bar{k})}}{\gamma\bar\mu}\right)^4\right]h^i_a=16\pi{G}S^i_a
	\label{equbojo}
\end{equation}
where $\eta$ is the conformal time defined by
$
d\eta=dt/a(t),
$	
$\bar\mu$ is a parameter related to the action of the fundamental Hamiltonian on a
lattice state that can be understood as the coordinate size of a loop whose
holonomy is used to quantize the Ashtekar curvature components, $n$ is so that $\bar \mu$ depends on
the triad component through $\bar\mu\sim\bar p ^n$, and $\gamma$ is the
Barbero-Immirzi parameter. The right-hand side of this differential equation 
corresponds to the source term of gravitational radiations. It also receives corrections from holonomies and 
vanishes in the absence of matter. The friction term and the last 
term of the left-hand side are given by the background evolution, as solved in the 
LQG framework. The background evolution of an isotropic and homogeneous universe 
with holonomy corrections has been solved
\cite{bojo2,bojo3,ashtekar1,ashtekar2} and is summarized in Bojowald \& Hossain \cite{bojo1}. With
this solution, one can compute the equation of motion for tensor
perturbation modes with $\bar\mu=\left({\bar{p}}/\lambda\right)^{n}$, 
$n\in[-1/2,0]$. The value of $n$ depends on the scheme adopted to quantize 
holonomies. Furthermore \cite{bojo1}, $\bar{p}=a^2(\eta)$ and 
$\lambda$ has to be chosen so that $\bar\mu$ has the dimension of a length.
The exact value of $\lambda$ and its dependence upon $n$ are still under debate. 
For the specific case $n=-1/2$, it was shown \cite{ashtekar2} that 
$\lambda=2\sqrt{3}\pi\gamma\ell^2_{\mathrm{Pl}}$ and it seems quite natural 
to  phenomenologically parametrize $\bar\mu$ by 
$
	\bar\mu\equiv\alpha\ell_{\mathrm{Pl}}\bar{p}^n.
$
In this case, the explicit value of $\alpha$ as a function of the 
Barbero-Immirzi parameter is known. Using of Eq. (29) to (31) from Bojowald \& Hossain
\cite{bojo1}, the above equation of motion can be re-written as a function of the
cosmological parameters (the scale factor $a(\eta)$ and the energy density
 of the background $\rho(\eta)$) and of three LQG parameters ($n$, $\alpha$ and
 $\gamma$)~:
\begin{equation}
	\left[\frac{\partial^2}{\partial\eta^2}+\frac{2}{a}\frac{\partial{a}}{\partial\eta}\frac{\partial}{\partial\eta}-\nabla^2-\left(\frac{2n\gamma^2\alpha}{M^2_{\mathrm{Pl}}}\right)\left(\frac{8\pi{G}\rho}{3}\right)^2a^{4+4n}\right]h^i_a=16\pi{G}S^i_a,
	\label{equ0}
\end{equation}
where we have replace $\ell_{\mathrm{Pl}}$ by $1/M_{\mathrm{Pl}}$. 
Introducing a new field $\Phi^i_a=a(\eta)h^i_a$, Eq. (\ref{equ0}) now reads~:
\begin{equation}
	\left[\frac{\partial^2}{\partial\eta^2}-\nabla^2-\frac{\ddot{a}}{a}-\left(\frac{2n\gamma^2\alpha}{M^2_{\mathrm{Pl}}}\right)\left(\frac{8\pi{G}\rho}{3}\right)^2a^{4+4n}\right]\Phi^i_a=16\pi{G}a(\eta)S^i_a,
	\label{equholo}
\end{equation}
where $\ddot{a}$ means second derivative according to the conformal time $\eta$.

This equation is easily interpreted if one remembers the dynamical 
equation for gravitons in a FLRW background without LQG corrections~:
\begin{displaymath}
	\left[\frac{\partial^2}{\partial\eta^2}-\nabla^2-\frac{\ddot{a}}{a}\right]\Phi^i_a=16\pi{G}a(\eta)\tilde{S}^i_a,
\end{displaymath}
with $\tilde{S}^i_a$ the source term in general relativity. Holonomy corrections
appear as a modification of the dispersion relation. This modification is 
encoded in the last term of the LHS of Eq. (\ref{equholo}) and depends on the 
dynamics of the universe through the scale factor, on its content through the 
energy density of the background and on the LQG parameters. It can also be
noticed that, in addition to its time dependence through the scale factor, the 
correction term scales, as expected, with
$
	E_{\mathrm{background}}/M_{\mathrm{Pl}}.
$

In this note, we focus on the influence of LQG on the tensor perturbations
of quantum origin during the inflationary phase and the source term will be 
set to zero.

For a de Sitter (dS) inflation, the scale factor and the
energy density of the background are given, as functions of the conformal time, 
by~:
\begin{displaymath}
	\begin{array}{rcl}
	
\displaystyle{a}(\eta)=-\displaystyle\frac{1}{H\eta}~~~{\mathrm{with}}~\eta<0~~~,~~~
		\displaystyle\rho(\eta)=\displaystyle\frac{3H^2}{8\pi{G}}.
	\end{array}
\end{displaymath}
Decomposing the field into its spatial Fourier modes, 
\begin{displaymath}
	\Phi(\vec{x})=\displaystyle\int\frac{d^3k}{(2\pi)^3}\phi_k\exp{(i\vec{k}\cdot\vec{x})},
\end{displaymath}
and using the two above formula in the graviton equation of motion with 
$n=-1/2$, one has to deal with the following Schr\"odinger-like equation~:
\begin{equation}
	\frac{\partial^2\phi_k}{\partial\eta^2}+\left(k^2-\frac{\nu}{\eta^2}\right)\phi_k=0,
	\label{equhalf}
\end{equation}
with
\begin{displaymath}
	\nu=2-\alpha\left(\frac{\gamma{H}}{M_{\mathrm{Pl}}}\right)^2
\end{displaymath}
where the triad indices have been dropped to lighten the writings. As $\alpha$ and $\gamma$ 
are of the order of unity, the holonomy corrections are roughly given by 
$H^2/M^2_{\mathrm{Pl}}$ (that is the ratio of the energy scale of inflation to 
the Planck scale). 

The resolution of Eq. (\ref{equhalf}) is straightforward \cite{martin,stegun} 
and is given by the linear combination of Bessel functions
\begin{equation}
	\phi_k(\eta)=\sqrt{-k\eta}\bigg(A_kJ_{\tilde\nu}(-k\eta)+B_kY_{\tilde\nu}(-k\eta)\bigg),
	\label{bessel}
\end{equation}
with $\tilde\nu=\sqrt{\nu+1/4}$ ($A_k$ and $B_k$ being two constants of 
integration determined by the initial conditions). Initial conditions are found 
by studying the region where the adiabatic vacuum can be defined \cite{martin},
which is possible because the squared frequency has the same behavior as in the 
standard inflationary case (only the value of $\nu$ is modified). Those 
initial conditions are completely equivalent to the requirement that the initial 
quantum state is the Minkowski vacuum. In the remote past, the adiabatic vacuum is given by plane wave 
solutions
\begin{equation}
	\lim_{k\eta\to-\infty}\phi_k(\eta)=\frac{4\sqrt{\pi}}{M_{\mathrm{Pl}}}\frac{e^{-ik\eta}}{\sqrt{2k}}.
	\label{ini-cond}
\end{equation}
Matching the general solution with the above asymptotic solution leads to~:
\begin{eqnarray}
A_k=\left(\frac{\pi\sqrt{2}}{M_{\mathrm{Pl}}\sqrt{k}}\right)\left(\frac{e^{i\beta}}{\cos{2\beta}}\right)~,~
B_k=\left(\frac{\pi\sqrt{2}}{M_{\mathrm{Pl}}\sqrt{k}}\right)\left(\frac{ie^{-i\beta}}{\cos{2\beta}}\right),
\end{eqnarray}
with $\beta=\tilde\nu\pi/2+\pi/4$. To compute the final power spectra, the 
general solution has to be expanded in the high amplification regime, 
$k\eta\to0$. In this regime,  $J_{\tilde\nu}$ tends to zero whereas
$Y_{\tilde\nu}$ diverges, encoding the behavior of the 
so-called growing mode. In the limit of high amplification at any values of $k$,
the tensor perturbation modes then behave as~:
\begin{equation}
	\lim_{k\eta\to0}\phi_k(\eta)\simeq{B}_k\frac{2^{\tilde\nu}\Gamma(\tilde\nu)}{\pi}\left(-k\eta\right)^{\frac{1}{2}-\tilde\nu}.
\end{equation}
From this, one recovers a power-law spectrum
\begin{equation}
	\mathcal{P}_{\mathrm{T}}(k)=A_{\mathrm{T}}k^{3-2\tilde\nu}~,~
	A_{\mathrm{T}}=\left(\frac{2^{1+\tilde\nu}\Gamma(\tilde\nu)H}{\pi{M}_{\mathrm{Pl}}\cos{(2\beta)}}\right)^2\left|\eta_f\right|^{3-2\tilde\nu}.
\end{equation}
The presence of holonomy corrections modifies the primordial power spectrum in 
two ways~: it changes its normalization and leads to a departure from scale 
invariance. One can easily check that setting $\gamma=0$, and consequently 
removing the LQG corrections, the standard scale invariant power spectrum 
($\tilde\nu=3/2$) is recovered.
\vspace{0.3cm}

Departure from scale invariance can be qualitatively inferred from 
Eq. (\ref{equhalf}). Amplification of a mode with wavenumber $k$ starts when 
this mode penetrates into the potential barrier. The amplification therefore 
starts for $\eta_c=-\sqrt{\nu}/k$. If the Barbero-Immirzi parameter is 
real-valued, then the critical time at which amplification starts is later
than in the general relativistic case and we expect the mode to be {\it less} 
amplified when LQG corrections are considered. Moreover, as the difference 
between the critical time with and the critical time without LQG corrections 
decreases for higher values of $k$, short wavelengths modes will be less affected 
than modes with longer wavelengths. Since the power spectrum is scale invariant 
in the standard approach, this LQG spectrum should become {\it blue}.

A Taylor expansion of the spectral of 
the primordial spectrum is given by~:
\begin{eqnarray}
	n_{\mathrm{T}}=3-2\sqrt{\nu+1/4} 
	\simeq\frac{2\alpha}{3}\left(\frac{\gamma{H}}{M_{\mathrm{Pl}}}\right)^2+\mathcal{O}(H^4/M^4_{\mathrm{Pl}}). \nonumber
\end{eqnarray}
The spectrum is indeed blue for a Barbero-Immirzi parameter whose value would,
{\it e.g.}, be inferred 
from the Bekenstein entropy of black holes \cite{carlo2}. 
It would become red for a pure imaginary Barbero-Immirzi parameter (as suggested
in the initial work from Ashtekar). However this case is highly disfavored --if
not forbidden-- in LQG at it leads to a non-compact gauge group.\\

Considering now the more realistic case of {\it slow roll} inflation, the $\nu$ 
parameter entering Eq. (\ref{equhalf}) is then given by~:
\begin{displaymath}
	\nu=\frac{1}{(1-\epsilon)^2}\left(1-\alpha\left(\frac{\gamma{H}}{M_{\mathrm{Pl}}}\right)^2\right)+\frac{1}{1-\epsilon},
\end{displaymath}
where $\epsilon=-\dot{H}/H^2$ is the first slow roll parameter. The computations
carried out in the dS inflation case leads, in this case (to the first order in LQG 
corrections), to~:
\begin{eqnarray}
	n_{\mathrm{T}}&\simeq&\frac{-2\epsilon}{1-\epsilon}+\frac{2\alpha}{(1-\epsilon)(3-\epsilon)}\left(\frac{\gamma{H}}{M_{\mathrm{Pl}}}\right)^2+\mathcal{O}(H^4/M^4_{\mathrm{Pl}}) \\
	&\simeq&-2\epsilon+\frac{2\alpha}{3}\left(\frac{\gamma{H}}{M_{\mathrm{Pl}}}\right)^2+\mathcal{O}(\epsilon{H}^2/M^2_{\mathrm{Pl}}). \nonumber
\end{eqnarray}
This underlines two contributions to the departure from scale invariance~: 
the slow-roll parameter contribution and the LQG contribution. If one admits a high value for the 
inflation energy scale, $H\sim10^{16}~GeV$, the L.Q.G. corrections to the tilt 
of the tensor spectral index is of the order of $10^{-6}$, which is well below 
the expected contribution from slow-roll parameters. However the $n=-1/2$ case 
is the worst one from the detection viewpoint. The general equation
of motion for the mode of wavenumber $k$ is given by~:
\begin{equation}
	\frac{\partial^2\phi_k}{\partial\eta^2}+\left(k^2-\frac{2}{\eta^2}-\frac{2n\alpha\gamma^2H^4}{M^2_{\mathrm{Pl}}(-H\eta)^{4+4n}}\right)\phi_k=0
\end{equation}
and clearly shows that, as far as $n>-1/2$, the LQG contribution will inevitably
play a crucial role when $\eta \rightarrow 0$. This will be studied elsewhere
\cite{grain},
together with inverse volume corrections. Finally it would be welcome to
consider this approach in the self-consistent LQG inflation
scenario \cite{bojoinflation}.


\section*{Acknowledgments}
We would like to thank M. Bojowald for vey helpful discussions.

\end{document}